\documentclass[aps,prb,preprint,showpacs,groupedaddress,superscriptaddress, amsmath,amssymb,longbibliography]{revtex4-2}

\usepackage[pdftex]{graphicx}

\usepackage{amsmath}
\usepackage{amssymb}
\usepackage{amsfonts}
\usepackage{dcolumn}% Align table columns on decimal point
\usepackage{bm}% bold math
\usepackage{color}
\usepackage{natbib}
\usepackage{float}
\begin{document}

%\preprint{}

\title{Ice phonon spectra and Bayes inference: a gateway to a new understanding of terahertz sound propagation in water} %Title of paper

\author{Alessio De Francesco}
    \email[Correspondence email address: ]{defrance@ill.fr}% Your name
    \affiliation{CNR-IOM \& INSIDE@ILL c/o Operative Group in Grenoble (OGG), F-38042 and Institut Laue Langevin, Grenoble, France}
\author{Luisa Scaccia}
   % \email[Correspondence email address: ]{defrance@ill.fr}% Your name
    \affiliation{Dipartimento di Economia e Diritto, Universit\`a di Macerata, Via Crescimbeni 20, 62100 Macerata, Italy}
\author{Ferdinando Formisano}
  % \email[Correspondence email address: ]{defrance@ill.fr}% Your name
    \affiliation{CNR-IOM \& INSIDE@ILL c/o Operative Group in Grenoble (OGG), F-38042 and Institut Laue Langevin, Grenoble, France}
\author{Eleonora Guarini}
    %\email[Correspondence email address: ]{defrance@ill.fr}% Your name
    \affiliation{Dipartimento di Fisica e Astronomia, Universit\`a di Firenze, via G. Sansone 1, I-50019 Sesto Fiorentino, Italy}
\author{Ubaldo Bafile}
    %\email[Correspondence email address: ]{defrance@ill.fr}% Your name
    \affiliation{Consiglio Nazionale delle Ricerche, Istituto di Fisica Applicata  ''Nello Carrara'', via Madonna del Piano 10, I-50019 Sesto Fiorentino, Italy}
\author{Ahmet Alatas}
    %\email[Correspondence email address: ]{defrance@ill.fr}% Your name
    \affiliation{Argonne National Laboratory, Advanced Photon Source, P.O. Box 5000 Upton, 11973 NY, USA}
\author{Scott T. Lynch}
    %\email[Correspondence email address: ]{defrance@ill.fr}% Your name
    \affiliation{Department of Physics, University of Wisconsin at Madison, 1150 University Avenue, Madison, WI, USA}
\author{Alessandro Cunsolo}
    %\email[Correspondence email address: ]{defrance@ill.fr}% Your name
    \affiliation{Department of Physics, University of Wisconsin at Madison, 1150 University Avenue, Madison, WI, USA}

%\date{\today}

\begin{abstract}

Understanding how molecules engage in collective motions in a liquid where a network of bonds exists has both fundamental and applied relevance. On the one hand, it can elucidate the ``ordering" role of long-range correlations in an otherwise strongly dissipative system; on the other hand, it can inspire new avenues to control such order to implement sound manipulation.
Water represents an ideal investigation case to unfold these general aspects and, across the decades, it has been the focus of thorough scrutiny. Despite this investigative effort, the spectrum of terahertz density fluctuations of water largely remains a puzzle for Condensed Matter physicists. To unravel it, we compare previous scattering measurements of water spectra with new ones on ice. Thanks to the unique asset of Bayesian inference, we draw a more detailed portrayal of the phonon response of ice. The comparison with the one of liquid water challenges the current understanding of density fluctuations in water, or more in general, of any networked liquid.

\end{abstract}

\maketitle %\maketitle must follow title, authors, abstract and \pacs

%\section{}
%\label{}
%\subsection{}
%\subsubsection{}

\section{Introduction}
\label{Intro}

It is a matter of common experience that liquids and solids oppose a different resistance to mechanical attempts to change their macroscopic shape. A possible way to test such a different rigidity is to induce a macroscopic mechanical perturbation generating the propagation of a density wave. An alternative pathway followed in inelastic scattering measurements consists in stimulating the propagation of density fluctuations at mesoscopic scales by irradiating a material with a beam of particle waves, such as neutrons or photons.
When dealing with a liquid system, such a density wave propagates with a strong damping and at macroscopic scales, uniquely in a direction parallel to the triggering force. In a structurally ordered solid such as a crystal, applied stresses propagate for a longer time and in directions both parallel and orthogonal to the applied force, in the form of longitudinal (LA) or transverse acoustic (TA) phonons, respectively.
However, this clear-cut distinction becomes more elusive over distances and time lapses respectively approaching the size of a single atom first neighbors cage and the period of such an atom's ``in cage" bounces.

Indeed, abundant experimental evidence endorses the conclusion that density waves in liquids bear strong evidence for terahertz viscoelasticity \cite{christensen2012theory}, i.e., they combine liquid-like and solid-like aspects \cite{ruocco1996equivalence,monaco1999viscoelastic,Scopigno_lithium2000,boon1991molecular}.
For instance, the transition between the viscous and the elastic regime in liquid water at about 280 K occurs at a few nm$^{-1}$, according to the combined results of Refs. [\onlinecite{santucci2006there}] and [\onlinecite{sette1996transition}]. 
Relevant points of the broad picture emerging from the intensive scrutiny of terahertz viscoelasticity in fluids can be summarized as follows:\\

1) The first spectacular manifestation of mesoscale viscoelasticity is the increase of sound velocity when probing distances and times roughly as small as those featuring first-neighbor atomic interactions. This trend reflects the enhanced rigidity of the system (elastic regime) at these so-called mesoscopic scales, typically much shorter than those covered by dissipative - diffusion or relaxation - processes in liquids.  \\

2) The transition to such an elastic regime is paralleled by an increased ability of the liquid to support the propagation of transverse acoustic waves.\\

3) As opposed to a long-standing belief, this viscoelasticity persists in thermodynamic domains extending well above the critical point \cite{gorelli2006liquidlike,simeoni2010widom}, although their actual delimitation raised some controversy \cite{bryk2017behavior,bolmatov2013thermodynamic,Yang2017,bolmatov2017emergent}.\\

4) Aside from its unsettled dependence on thermodynamic conditions, high-frequency viscoelasticity also has a still poorly understood link with the nature of microscopic interaction; for instance, it is unclear why it is substantially more pronounced in specific relatively low-viscosity systems as water \cite{cunsolo2015thz,cunsolo1999experimental,monaco1999viscoelastic}, carbon dioxide \cite{Sampoli2009}, deuterium \cite{guarini_d2}, and noble gases \cite{de1983short,van1985density,cunsolo2001microscopic}, while it is often barely detectable in liquid metals \cite{balucani,balucani1993liquid,guarini2013,guarini_ag,def2021Silver}.\\

Far from being a mere academic endeavor, gaining insight on these topics could deepen our understanding of subjects as fundamental as the very nature of liquid aggregation. Also, it will likewise disclose new avenues in the emerging domain of sound propagation design and engineering, a field whose enormous practical interest has been recognized \cite{maldovan2013sound}, yet not fully explored.

On the experimental side, terahertz phonon propagation in condensed materials can be directly probed by inelastic scattering methods, such as Inelastic X-Ray (IXS \cite{cunsolo2021thz,sinha2001theory}) and Neutron Scattering (INS, \cite{squires1996introduction,lovesey1984theory}).
Conceptually, IXS or INS spectroscopic probes resemble large microscopes ``pointed on the dynamics", which can be zoomed in to focus on dynamic events occurring over various scales. This goal is ordinarily achieved by suitable tuning of the two main variables of the scattering event, namely energy and momentum exchanged between the probe wave particles (here assumed to be photons), respectively referred to as $E = \hbar\omega$, and $\bf{p}=\hbar\bf{Q}$. Here, $\hbar$ is the reduced Planck constant, while $\bf{Q}$ ($\omega$) represents the exchanged wave-vector (frequency), i.e., the difference between before-scattering and after-scattering photon wave-vector (frequency).
For small enough values of $Q=\vert\bf{Q}\vert$ and $E$, the target system appears as a continuum whose dynamic response is probed as an average over many microscopic events.
Conversely, at extremely large $(Q,E)$'s, distances and times spanned become so small that the only event observed is the free recoil of the single struck atom after the collision with the photon and before subsequent first-neighbor interactions.
Clearly, all dynamic processes characterizing the interatomic dynamics of the sample can be captured, somewhere between these two limits, by adequately tuning the spectroscopic probe.
Not surprisingly, the developments of the scattering techniques directly mapping this broad dynamic domain, INS and IXS, have dramatically improved the current understanding of high-frequency viscoelasticity phenomena in an impressively varied family of disordered systems.
%\subsection{Unveiling terahertz sound propagation in %water}\label{water}
Although water is one of the disordered systems most thoroughly studied across various decades, its deep scrutiny has often evidenced seemingly anomalous and poorly understood behaviors, albeit sometimes later recognized as ``normal effects" that water shares with many other liquids.
Notable examples include the high propagation speed of the terahertz sound mode, once called ``fast sound" \cite{santucci2006there,sette1996transition}, and the onset of a shear mode propagation \cite{cunsolo2012transverse,cunsolo2016,cimatoribus2010mixed,pontecorvo2005high,sampoli1997mixing}.
Although similar phenomena are being reported for an increasing number of disordered systems, the case of water found a straightforward and broadly accepted explanation in terms of hydrogen bond (HB) dynamics.
According to this interpretative scheme, \cite{walrafen1996low,walrafen1964raman} high frequency transverse and longitudinal acoustic waves in water respectively couple with the bending of two HBs linking triplets of oxygen atoms, or to the stretching of the HBs connecting two adjacent oxygens.
Despite the remarkable insight achieved over the decades, further experimental effort is still needed to clarify a few more subtle aspects of water dynamics.
These include the microscopic mechanism leading shear acoustic modes of water to exhibit a nearly flat $Q$ dependence at the crossover between quasi-macroscopic and mesoscopic distances, while gradually becoming predominant over the longitudinal acoustic modes.

From a merely experimental perspective, a natural pathway to understand better the collective movements of water molecules is to observe how they compare with those in the solid. Indeed, the study of the acoustic response of  (polycrystalline) ice presents significant simplifications: 1) both atomic diffusion and HB network relaxations are frozen, thus yielding no visible contribution to sound propagation; 2) the lifetime of acoustic modes is much longer, and the related spectral features correspondingly sharper.
Nonetheless, as illustrated in the remainder of this paper, the phonon response of ice is far from being trivially interpreted and characterized. In this paper, we discuss an experimental attempt to elucidate similarities and distinctive behaviors of the terahertz dynamics of ice and water by comparing their IXS spectra.
Specifically, we jointly investigate collective excitations in water and phonon modes in polycrystalline hexagonal ice (Ih). On a rigorous ground, inelastic excitations in the spectrum of a polycrystal should not be referred to as phonons owing to their projection along multiple crystallographic directions. However, in the following we will resort to this broadly used nomenclature for consistency with existing literature.

The use of a Bayesian inference-based modeling of IXS lineshapes reveals an unprecedentedly complex phonon behavior of ice, while rectifying and complementing the current understanding of terahertz acoustic excitations in liquid water.

\section{The measurement}\label{Meas}

\begin{figure}
	\centering
	\includegraphics[width=\linewidth]{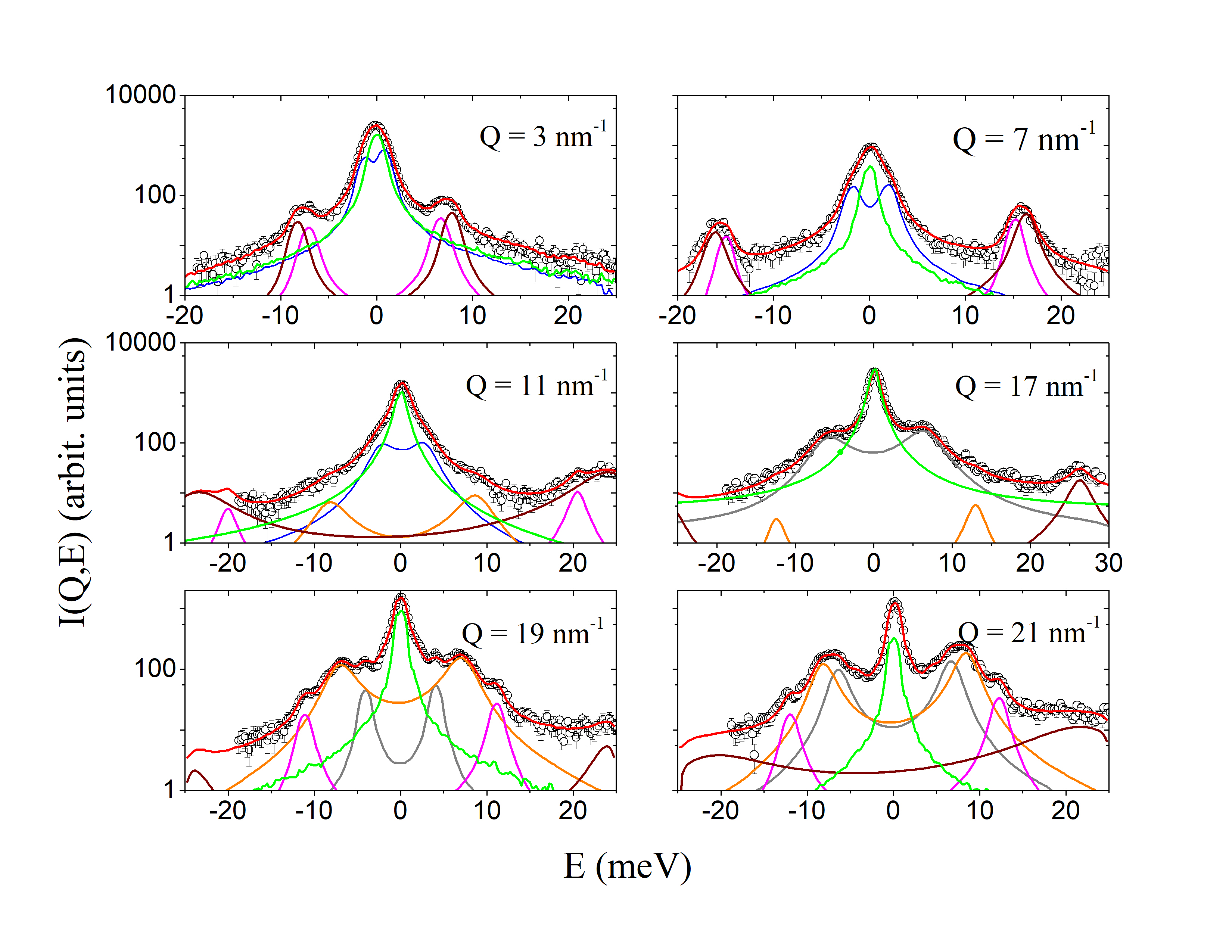}
	\caption{\textbf{Inelastic X-Ray Scattering (IXS) spectra of polycrystalline ice and their model lineshapes.} IXS spectra measured at representative $Q$'s and $T = 225$ K discussed in this work (open circles) compared with corresponding best-fit lineshapes (red lines through data) along with their inelastic DHO model components (see text): the low frequency transverse mode (blue line), the two intermediate frequency optic-like modes (grey and orange lines), the longitudinal acoustic mode (magenta) and a higher frequency mode (wine line). The green line represents the instrumental energy resolution. Error bars are estimated as the square root of the scattering intensity counts.}
	\label{Fig1}
\end{figure}

Measurements were executed at the Sector 30 beamline \cite{said2011new,toellner2011six} of the Advanced Photon Source at Argonne National Laboratory. The instrument was operated using the $\approx$ 23.7 keV harmonic of the undulator source, which corresponds to the Si(12~12~12) backscattering reflection from both the monochromator and energy analyzers. Spectral acquisitions covered the 3 $\div$ 21 nm$^{-1}$ $Q$ range with a 2 nm$^{-1}$ step.
The instrumental resolution function was measured through the IXS signal from a Plexiglas sample at the $Q$ of its first sharp diffraction maximum, i.e., about 10 nm$^{-1}$.  The resulting spectrum had a 0.8 meV broad (half width at half maximum, HWHM) nearly Lorentzian profile, sufficient to properly resolve the relevant spectral features discussed in the remainder of this paper. All measurements were executed at a temperature of 225 K. 
The polycrystalline nature of the sample makes its global orientation irrelevant. As a consequence, the measurement probed the constant-$Q$ phonon response of ice as an average over the various crystallographic directions.
Fig. \ref{Fig1} illustrates the emergence of phonon excitations in the IXS spectra of ice by comparing raw and best-fitting model spectra for selected $Q$ values; the semilogarithmic plot enables a full appreciation of relevant lineshape features across various intensity decades.
Model lineshapes, obtained as discussed in Sec. I of Supplemental Materials, consist of a $\delta$-function describing the elastic scattering and individual DHO terms accounting for inelastic excitations in the spectrum.  The most plausible number of DHO components was determined \textit{a posteriori}, based on the outcome of the Bayesian inferential analysis of measured data, as described in our previous works \cite{de2016bayesian,de2018damping,de2019model,de2020shaping,de2020onset}. Individual DHO components of the best-fitting total spectral distribution are also included in Fig. \ref{Fig1} for reference. The assignment of each spectral contribution is summarized in the caption of Fig. \ref{Fig1} and illustrated in Table I.

\begin{table}[h!]
\begin{center}
\begin{minipage}{274pt}
\caption{Polycrystalline Ice modes assignment}\label{tab1}%
\begin{tabular}{@{}lllll@{}}
\toprule
Mode & Label & Assignment & &Symbol color \\
%\midrule
 1    & TA  & Transverse Acoustic  & &blue  \\
 2    & O$_L$   & Lower Optical mode  & & grey  \\
 3    & O$_H$   & Higher Optical mode & &orange  \\
 4    & LA& Longitudinal acoustic mode of center of mass  &   & magenta  \\
 5    & HFM   & highest frequency steeply dispersive mode & & wine \\
\botrule
\end{tabular}

\end{minipage}
\end{center}
\end{table}

Notice that the same color code is maintained consistent throughout all graphs included in this work.
The IXS spectra in Fig. \ref{Fig1} are reported in the region where they have been modeled. This region, varying in extent and symmetry with respect to the central peak position, essentially coincided with the energy range covered by the corresponding energy resolution measurement.

One can readily notice that spectral shapes displayed in Fig. \ref{Fig1} are dominated by multiple inelastic features, which, in some cases, partially overlap giving rise to inelastic peaks visibly broader than the resolution function, as particularly evident in the $Q$ = 21 nm$^{-1}$ spectrum. Overall, Fig. \ref{Fig1} provides a comprehensive rendering of the complexity of the phonon response of polycrystalline ice down to mesoscopic distances. The $Q$-evolution of various phonon modes will be discussed in further detail in the next Section, here we only stress that these modes display quite distinctive $Q$ dependencies revealing their either acoustical or optical origin. 

\section{Discussion of results}\label{Disc}
\begin{figure}
	\centering
	\includegraphics[trim={0 1cm 0 0},width=.6\linewidth]{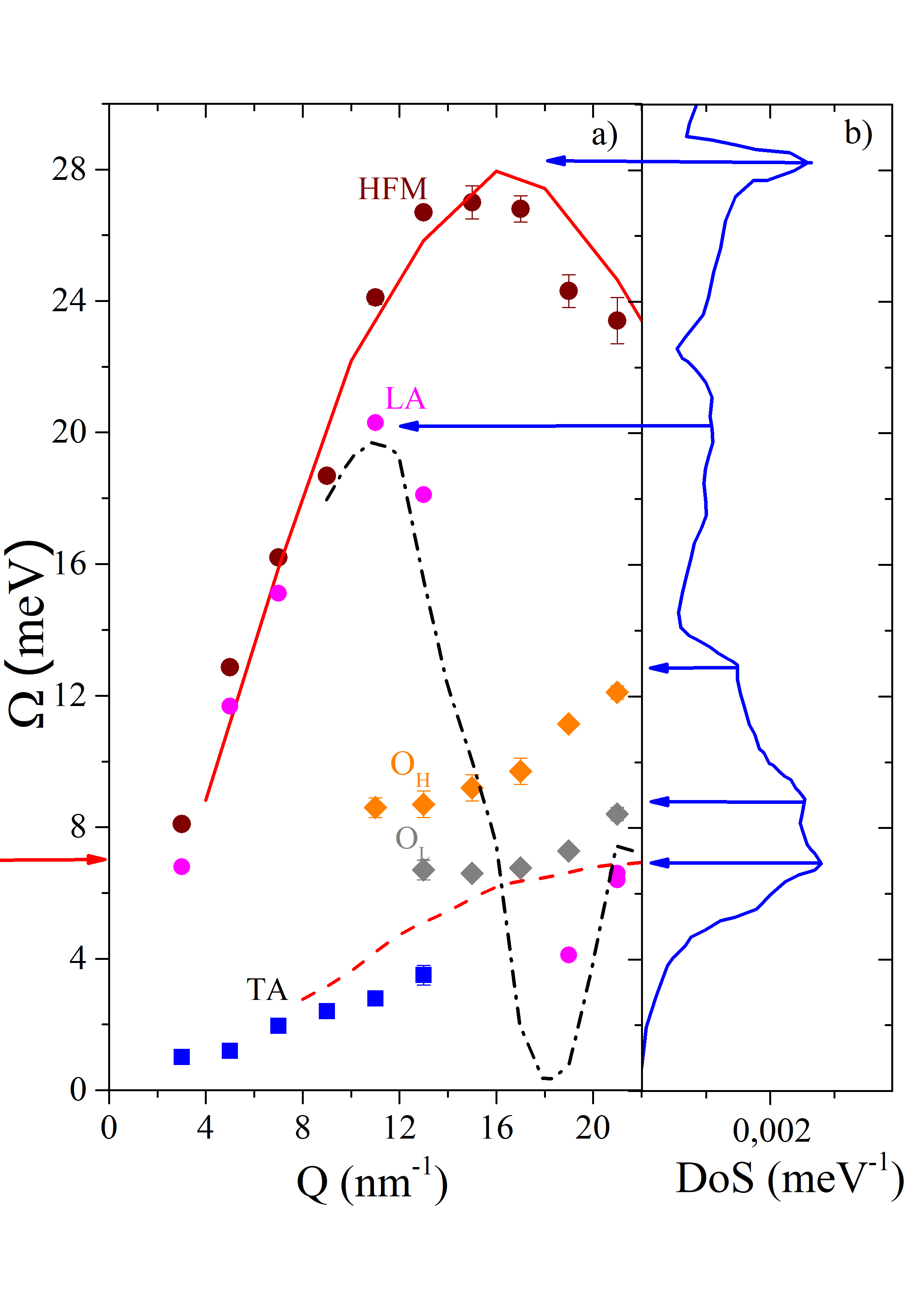}
	\caption{\textbf{Phonon branches of polycrystalline ice and corresponding sound dispersions in liquid water.} \textbf{Panel a}: The plot displays the following dispersion branches: the transverse acoustic (TA) mode (blue squares); the longitudinal acoustic (LA) mode (magenta dots); the high-frequency mode (HFM, wine dots); the lower and higher energy optical modes O$_L$ and O$_H$ (grey and orange lozenges, respectively. Error bars in the dispersion curves are estimated through the standard deviations of the corresponding posterior distribution functions. The lines refer to literature results and are colored in black or in red for ice and liquid water respectively. Specifically, they represent the longitudinal mode of liquid water at 263 K and 2 kbar \cite{pontecorvo2005high} (solid red line); the transverse mode of room temperature  D$_2$O \cite{cunsolo2012transverse} (dashed red line) and the LA mode of polycrystalline ice Ref. \cite{criado1993phonon} (black dash-dotted line). The horizontal red arrow points to the optical frequency measured by optical spectroscopy \cite{mazzacurati1981interaction}. \textbf{Panel b}: The phonon Density of States measured by neutron scattering on single hexagonal ice (\cite{del2021density}, blue line). Horizontal arrows are guides to the eye showing the rough correspondence between DoS features and dispersion curves extrema.
	}
	\label{Fig2}
\end{figure}

The good consistency between the model and measured lineshapes clearly emerges from the Fig. \ref{Fig1}. Selected IXS spectra are compared with their best-fitting model lineshapes discussed in the  Sec. I  of Supplemental Materials.
Overall, phonon spectra presented in this paper are qualitatively similar to those reported in the IXS measurement described in Ref. \cite{ruocco1996equivalence}. However, the latter was performed on a polycrystalline sample seemingly more ordered than the present one, as suggested by the weaker elastic contribution, yet with a resolution significantly coarser, as to be expected considering the still pioneering stage of high-resolution IXS.
Once assessed the ability of the model to reproduce the detail of the spectral shape, we can now focus on best-fit values of phonon frequencies, whose $Q$ dependence and assignment are included in  Fig. \ref{Fig2}. 

%The same figure also displays for reference the dispersion %curves from Ref. [\citenum{de2020terahertz}].
 
As apparent from the dispersion curves displayed in Fig. \ref{Fig2}, the low-energy ($<$ 25 meV) phonons of ice probed by this measurement show some rather peculiar features, previously either undetected \cite{ruocco1996equivalence,renker1969phonon} or only partially reported by studies in the literature \cite{cimatoribus2010mixed,criado1993phonon}. Specifically:

1) The longitudinal acoustic phonon of ice, i.e., the one characterized by the sharpest (linear) low $Q$ growth in Fig. \ref{Fig2}a, actually splits into two branches, consistent with what was reported in a previous IXS work by Cimatoribus and collaborators \cite{cimatoribus2010mixed}, and previous lattice calculations \cite{Johansson_Ice1996}. 
Most importantly, this trend might suggest reconsidering the surprising line broadening of the longitudinal acoustic mode of both water and polycrystalline ice reported in Ref. \citenum{ruocco1996line}. The resolution limitation of such a pioneering IXS measurement is the most likely explanation for the mode-splitting appearing as a line-broadening.
At low enough $Q$'s, the two split branches attain similar values following a parallel $Q$-dependence while they spread apart at larger $Q$'s. To estimate the low wavevector propagation speed of these two modes, say  ${c_1}$ and ${c_2}$, we looked at the slope of the three lowest $Q$ points of the corresponding phonon branches (Fig. \ref{Fig2}). We found that the ratio of the sound speeds ${c_1}/{c_2}$ = 1.09 $\pm $ 0.07, a value consistent within the error to that of $\sqrt{M_{CM}/M_{O}}=1.06$ where $M_{CM}$ and  $M_{O}$ are the masses of the water molecule and the oxygen atom, respectively. This finding might suggest that, in the linear dispersion regime, the modes reported in Fig. \ref{Fig2} as LA and HFM, might be connected to collective vibrations involving either the whole H$_{2}$O molecule or the oxygen atoms only. This is the main rationale behind the identification of the lower frequency component of the doublet to the ordinary longitudinal acoustic (LA) mode.

2) From moderate to high $Q$ values, the LA branch acquires a negative $Q$ slope, bending downward to some minimum at about 17 nm$^{-1}$, in agreement with what was previously observed by the molecular dynamic simulation by Criado and collaborators \cite{criado1993phonon}. Conversely, at about the same $Q$ value, the HFM reaches its maximum at about 25 meV. Noticeably the plot suggests that the HFM branch is the one most closely resembling the longitudinal acoustic dispersion of liquid water, which is also reported in Fig. \ref{Fig2}a, as measured by IXS at 263 K and 2 kbar \cite{pontecorvo2005high}.

4) Noticeably, two additional weakly dispersing modes show up in the spectrum inside the 7-12 meV, $Q>$ 10 nm$^{-1}$ window: a lower and a higher optical mode here labeled as O$_L$ and O$_H$, respectively. We assign an optic origin to these branches owing to their mild $Q$ dependence and seemingly non-vanishing low $Q$ trend, which distinguishes them from the transverse acoustic mode, sitting at an even lower frequency. The existence of multiple phonon branches at similar energies was reported in a computational study  by Criado and collaborators on polycrystalline ice \cite{criado1993phonon}  and by previous works in the single crystal at ambient \cite{renker1969phonon}, or higher pressures \cite{strassle2004phonon}. Also, present results are consistent with Density of State measurements by incoherent INS \cite{li1991inelastic,Klug_PRB_1991} and optical spectroscopy \cite{mazzacurati1981interaction}. In Fig. \ref{Fig2}a, a horizontal arrow points to the $Q$ = 0 (energy axis) value corresponding to one of the optical mode energies reported by the latter work. Revisited \textit{a posteriori} in the light of current results, the optical-like behavior of medium $Q$ phonons of ice is also consistent with the findings of coherent scattering measurements by INS  \cite{renker1969phonon,cunsolo2012transverse} and IXS \cite{ruocco1996equivalence,Schober2000}.

Fig. \ref{Fig2}a suggests that a proper discernment of two distinct modes, LA and HFM, may be challenging at some $Q$ values. However, the inference of a mode doublet with such a narrow energy separation is pondered against a precise estimation of a joint probability distribution. The involved energies are so close that the modes partially overlap, giving origin to a single peak yet still significantly broader than the instrumental energy resolution and somewhat flat atop as expected for an excitation composed by multiple spectral modes of similar amplitude. According to the probability rating of the Bayesian algorithm, the most plausible number of DHO modes participating to such an excitation is two.
It would be tempting to compare the inelastic modes of the considered polycrystal to its single-crystal counterpart thoroughly studied in the literature \cite{renker1969phonon,Wehinger_2014}. However, this comparison would be problematic since, as mentioned, excitations of our sample are averaged over the various crystallographic directions as opposed to the single-crystal case. For instance, the HFM dispersion curve might be contaminated by higher energy optical modes especially in the apex region. In this perspective, the single-crystal phonon Density of States (DoS), being a wavevector-averaged (non-directional) property, provides a better quantity to compare with. For reference, we included this spectral function in Fig. \ref{Fig2}b as measured by neutron scattering on single crystal Ih \cite{del2021density}. The horizontal arrows serve as guides to the eye, highlighting the overall consistency between relevant DoS features and the extremes of dispersion curves derived in the current measurement.

Looking at the sound dispersions drawn in Fig. \ref{Fig2}a, one may notice the disappearance of the transverse acoustic (TA) branch for $Q >$ 14 nm$^{-1}$. We tend to rule out any physical rationale behind this trend, instead blaming the tendency of several inelastic features to cram into a narrow energy window, thus challenging the algorithm's ability to distinguish them, especially those having a faint spectral fingerprint. In fact at $Q =$21 nm$^{-1}$ there might be a hint for the presence of the transverse mode (see Fig. S1 and S2 of the SM) but, in our opinion, not with enough experimental evidence. Very likely at this $Q$ this mode is still overwhelmed by the near more intense optic modes which makes the detection uncertain.

\subsection{Comparing polycrystalline ice and liquid water}

\begin{figure}
	\centering
	\includegraphics[width = .7\textwidth]{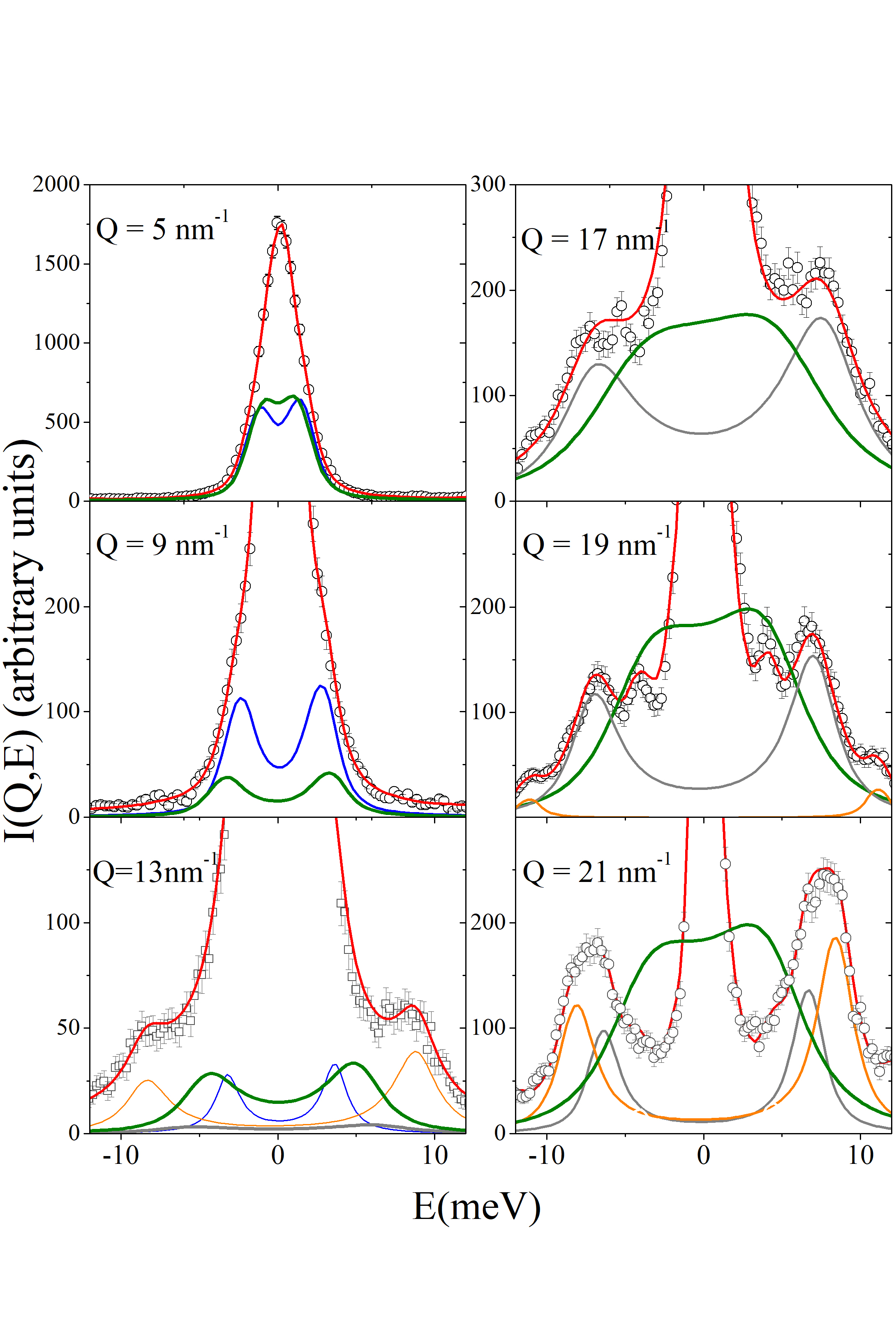}
	\caption{\textbf{Comparison of low and intermediate frequency phonon modes in polycrystalline ice and in pure water.} Selected spectra measured in this work at the indicated $Q$s are here displayed in expanded energy transfer and intensity windows. Experimental spectra in ice (black circles) are compared with best-fitting model profiles (red line), the transverse low energy component (blue line), and the two optical-like ones (grey and orange lines), whenever present. The thick olive green line represents the low energy mode in pure water at 300 K as obtained through the lineshape analysis discussed in Ref. [\citenum{de2020terahertz}].}
	\label{Fig3}
\end{figure}

Fig. \ref{Fig3} compares, in a more restricted intensity and energy window, the phonon lineshapes of ice with the low energy mode we measured in a previous work on liquid water \cite{de2020terahertz}. This comparison becomes especially informative when considered in combination with results in Fig. \ref{Fig2}.

Before digging further into this aspect, we recall here that the emergence of the low energy mode in the spectrum of water has been the focus of intensive experimental \cite{cunsolo2012transverse,cunsolo2016,cimatoribus2010mixed,pontecorvo2005high,sampoli1997mixing}, and computational \cite{sampoli1997mixing,Balucani_1996} studies, which, despite some controversy \cite{sacchetti2004brillouin,petrillo2000high}, endorsed the now broadly accepted assignment to a transverse acoustic excitation.
However, results in Fig. \ref{Fig2} and  Fig. \ref{Fig3} call for a reconsideration of this assignment, which, ironically enough, was originally inspired right by the comparison with the ice spectrum \cite{sette1996transition,ruocco1996equivalence}. 
Indeed, Fig. \ref{Fig2} suggests that the transverse mode of water does coincide with the TA branch of ice at the lowest $Q$'s only. Conversely, upon $Q$-increase, it seems to mimic the lower energy  optical phonon of ice, while acquiring, in turn,  a weaker $Q$ dependence.
This conclusion is further endorsed by Fig. \ref{Fig3}, which shows that the low-energy spectral feature of water closely resembles its transverse acoustic counterpart in ice for $Q \leq$ 9 nm$^ {-1}$ only. Conversely, at larger $Q$'s, it rather parallels the optical phonons dominating the ice spectrum, even though the latter, owing to the lack of damping, have visibly higher inelastic shift. These evidences  urge us to infer that collective modes in water correspondingly acquire a dominating optic-like character in this $Q$ region.
On a more general perspective, the above arguments might seem to question the very nature of a collective mode in a fluid, as they imply that such a mode can actually combine together independent phonon vibrations. Moreover, the damping enhancement stemming from diffusive and relaxation processes can make even fuzzier the physical interpretation of the inelastic mode of a liquid. However, the scenario is not as bewildering for water, at least when its spectrum is compared to that of its solid-state, ordered, counterpart. In fact, although vibrational modes of different nature seem to participate in the low energy mode of water, they dominate complementary $Q$ windows: optical modes become preponderant upon $Q$ increase while (transverse) acoustic ones follow just the opposite trend. From this perspective, one is authorized to conclude that, upon $Q$ increase, the low frequency bump in the water spectrum changes its dominant character from primarily acoustic to mainly optical.

We believe that this finding is of high relevance since, thus far, no experimental or computational evidence has ever challenged the consensual assignment of such a spectral feature to an acoustic phonon-like (transverse) mode.
One of the most visible signs of this acoustic-to-optic transformation is the gradually  vanishing group velocity $v_g =\partial \Omega(Q)/\partial Q$ (with $\Omega(Q)$ being the phonon frequency), which reveals an underlying mode localization. This conclusion rests on the assumption that similar vibration modes are shared by water and ice, although only in ice they can be properly resolved due to their sharp profile and relatively tiny energy offset. 
We finally notice that, even in ice, in some $Q$ window, low energy modes tend to overcrowd the quasielastic region of the spectrum, thus challenging the ability of the Bayesian algorithm to properly identify their presence. This ``mode overcrowding" combined with the overwhelming intensity of adjacent optical modes, is likely, as mentioned, the reason of an apparent disappearance at a certain $Q$ value of the TA phonon (see Fig. \ref{Fig2}a).

\begin{figure}
	\centering
	\includegraphics[width=0.7\linewidth]{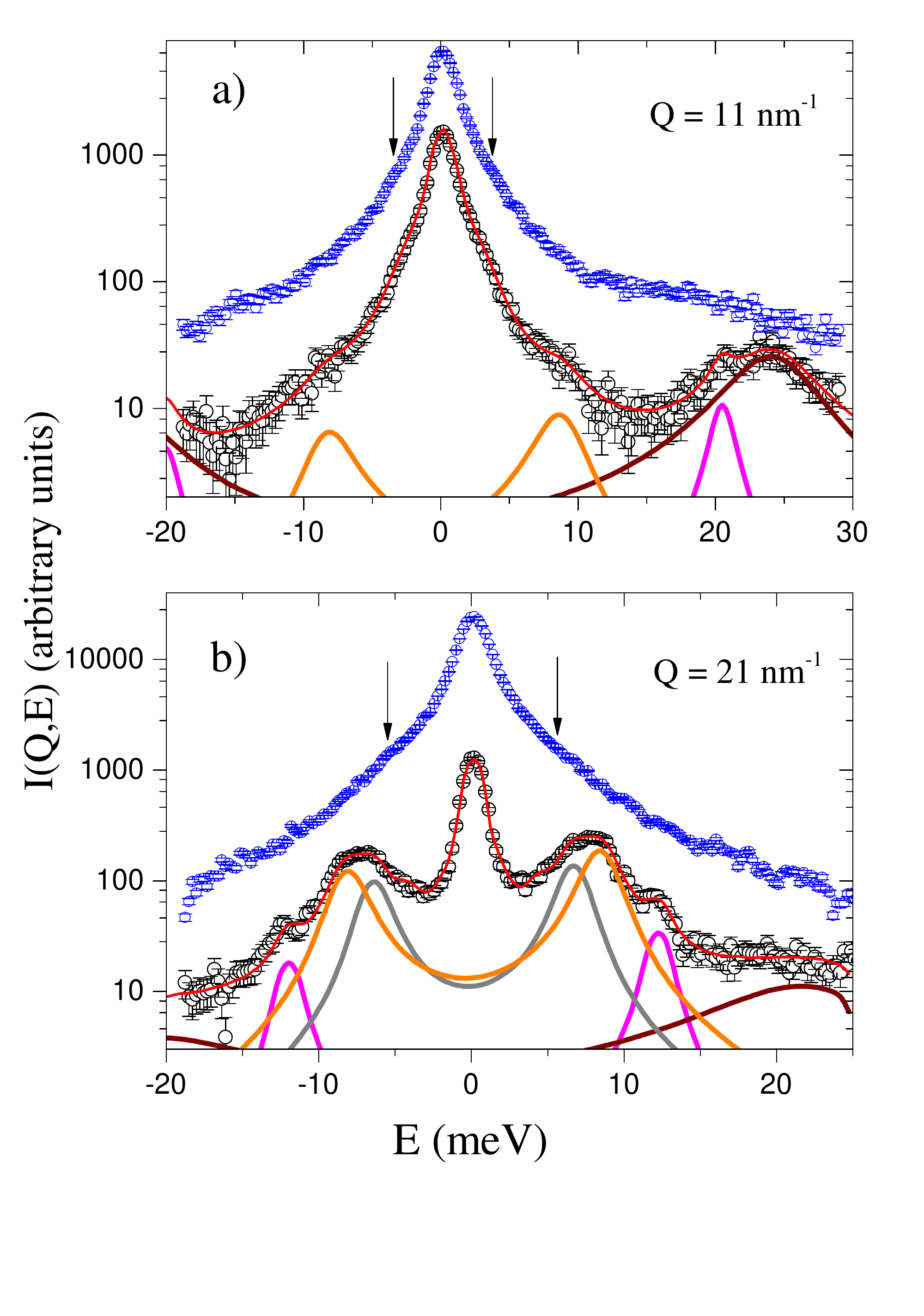}
	\caption{\textbf{High-frequency dynamics in polycrystalline ice and in water.} IXS spectra (black circles) measured at two selected $Q$ values are reported along with corresponding best-fitting curves (red lines) and both the HFM (wine line) and the LA (magenta line) energy model (DHO) components. With the orange and grey lines the optical-like modes are reported as in the previous figures. The corresponding IXS spectra of pure water measured in Ref. [\citenum{de2020terahertz}] are also shown (blue circles) after arbitrary vertical shift.  The two vertical arrows point at the position of the transverse mode energy in pure water.}
	\label{Fig4}
\end{figure}

Let us now comment on the comparison between current measurements on ice and those on liquid water reported in Ref. [\citenum{de2020terahertz}] and also included in Fig. \ref{Fig4} after arbitrary vertical shift. Both the HFM mode of ice and its counterpart in water weaken upon $Q$-increase, down to almost vanish at high $Q$'s. In ice the two optical branches, O$_L$ and O$_H$, essentially take it over in the inelastic wings. Notice that the naked eye's perception of this mode in the 21 nm$^{-1}$ spectrum of ice largely owes to the semilogarithmic representation. This strong suppression of the HFM is also witnessed by the rapid increase of the damping parameter with increasing $Q$ (Fig. S4 in the Supplemental Materials). In water, such an effect is even more pronounced, due to the enhanced damping which makes this mode barely distinguishable (if at all) from the large inelastic wings of the dominating transverse mode, whose position is roughly indicated by the two vertical arrows (see also Fig. 2 of Ref. [\citenum{de2020terahertz}]). 
Although the naked eye can barely discern the HFM mode in the  21 nm$^{-1}$ spectrum of Fig. \ref{Fig4}, the Bayesian algorithm endorsed without ambiguity its presence. One can quantify the statistical significance of this excitation through the high probability rating of a lineshape model explicitly containing it (66 $\%$). Nonetheless, the posterior of the related excitation frequency (see Fig. S2 of the Supplemental Information) is quite broad and weak, thus revealing the uncertain location of this otherwise genuine phonon excitation. Of course, the significance of a poorly discernible spectral feature should be pondered against all diagnostic factors available, mainly two in the present case: the probabilistic support of the Bayesian inference and the continuity with the trend followed by neighboring Q-points. Both indicators persuaded us about  the presence of a loose HFM feature at about 23 meV in the $Q$ = 21 nm$^{-1}$ spectrum.

\subsection{The key role of Bayesian analysis}

At this stage, it is crucial to understand how the reported result fit into the broader context of available literature data, and, in particular, why the outcome of this study brings to the surface a more complex phonon behavior than previously observed by individual works.  
Aside from some improvement in the statistical accuracy in the currently measured spectra, the use of Bayesian analysis here provides the critical asset of a minimally biased modelling of measured spectra. In this approach, the model itself is entrusted to establish on a probabilistic basis the most plausible number of phonon modes in the measured spectrum. 
This provides a natural protection against two somewhat opposite risks: the confirmation bias, holding off the discovery of new spectral excitations, and the model over parametrization, as would be the inclusion of too many independent excitations in the lineshape model. Bayesian inference endows researchers with these antidotes also thanks to the intrinsic implementation of the Occam razor principle, or "\textit{lex parsimoniae}", prescribing that, among equally plausible competing models, the simplest, i.e., the one with a smaller number of free parameters, is always to be privileged \cite{MacKay03}.
Obviously, this aspect can become a game changer when facing inherent ambiguities of a measured spectral line, perhaps exacerbated by a  less than adequate statistical accuracy. As an example we can consider the observation of an anomalously wide spectral line, to be alternatively interpreted as a genuine line broadening, secondary to an increased excitation lifetime, or the emergence of a new line, indicating the onset of a mode-splitting. Of course, Bayesian analysis only identifies the probabilistically most grounded hypothesis based on the measurement obtained, yet it cannot substitute investigators in difficult assessment of its physical plausibility. Further details on the Bayesian inference method, model choice and results can be found in the Supplemental Materials , specifically in Figs. S1, S2, S3 and S5.

\section{Conclusion}\label{sec3}
We have presented an Inelastic X-Ray Scattering study of the terahertz dynamic response of water, in which new  measurements in the solid phase are compared to our previous results in the liquid. In this work, a Bayesian inference based analysis of measured spectra represented a pivotal tool to unveil the full complexity of the phonon response of ice, thereby improving, via direct comparison, our understanding of the water's one. 

In summary, our data show that, on approaching microscopic scales, high energy modes becomes increasingly impeded. This hindrance is reflected, on one side, by the rapid increase of the longitudinal acoustic damping of the highest frequency mode, and, on the other, by a relative attenuation compared to optic modes which gradually become dominant in adjacent spectral windows. Indeed, by observing how measured spectra and dispersion curves derived from them compare in liquid and solid phases, we are urged to conclude that optic terahertz modes dominate the spectrum of density fluctuations of both water and ice at sufficiently short distances. This finding is especially noteworthy as clearly at variance with our long-lasting assumption of a dominating acoustic character of collective modes of water.
Despite this evidence, our results also stimulate the scientific community with new interpretative challenges. Likewise, the most urgent one is whether the observed behavior is distinctive or shared with other networked fluids. A dedicated experimental and computational effort is planned to elucidate this aspect hopefully improving our understanding of the high-frequency response of disordered systems.

\begin{acknowledgments}

This research was funded by Ministero dell'Istruzione dell'Universit\`a e della Ricerca Italiano (Grant No. PRIN2017-2017Z55KCW).

\end{acknowledgments}


%\bibliography{your-bib-file}

 %``â€ 
%\bibliographystyle{plain}
\bibliography{biblio_ice}

%%%%%%%%%% Merge with supplemental materials %%%%%%%%%%
\pagebreak
\widetext
\begin{center}
\textbf{\large Supplemental Materials: Ice phonon spectra and Bayes inference: a gateway to a new understanding of terahertz sound propagation in water}
\end{center}

%%%%%%%%%% Merge with supplemental materials %%%%%%%%%%
%%%%%%%%%% Prefix a "S" to all equations, figures, tables and reset the counter %%%%%%%%%%
\setcounter{equation}{0}
\setcounter{figure}{0}
\setcounter{table}{0}
\setcounter{section}{0}
\setcounter{page}{1}
\makeatletter
\renewcommand{\theequation}{S\arabic{equation}}
\renewcommand{\thefigure}{S\arabic{figure}}
\renewcommand{\thetable}{S\arabic{table}}
%\renewcommand{\bibnumfmt}[1]{[S#1]}
%\renewcommand{\citenumfont}[1]{S#1}
%%%%%%%%%% Prefix a "S" to all equations, figures, tables and reset the counter %%%%%%%%%%
\section{Experimental details and Bayesian approach}\label{methods}

To adequately reproduce the highly structured spectrum of density fluctuations of ice, i.e., the dynamic structure factor $S(Q,E)$, we considered minimally invasive hypothesis on its shape, which was assumed to include the superposition of an unknown number of elementary excitations. Namely:

\begin{eqnarray}
S(Q,E)&=&A_{e}(Q)\delta(E)+[n(E)+1]\frac{E}{k_BT}\Bigg\{\sum_{j=1}^{k}A_{j}(Q){\rm DHO}_j(Q,E)\Bigg\},
\label{SqE}
\end{eqnarray}

\noindent here $\delta(E)$ is the Dirac delta function of $E$ describing the elastic response and having integral $A_e$ and $n(E)=(e^{E/k_{B}T}-1)^{-1}$ is the Bose statistics factor accounting for the detailed balance condition. The $k$ phonon excitations contributing to the spectrum are approximated by Damped Harmonic Oscillator (${\rm DHO}_j(Q,E)$) multiplied by  intensity factors $A_j(Q)$ where:

\begin{eqnarray}
\noindent {\rm DHO}_j(Q,E)=\frac{2}{\pi}\frac{\Omega_{j}^{2}(Q)\Gamma_{j}(Q)}
{(E^2-\Omega_{j}^2(Q))^2+4[E\Gamma_{j}(Q)]^2},
\label{DHO}
\end{eqnarray}

\noindent and where $\Omega_{j}(Q)$ and $\Gamma_{j}(Q)$ are the undamped energy and the damping coefficient of the \emph{j}th DHO excitation. We stress again that the number $k$ of the ${\rm DHO}_j(Q,E)$ excitations and their shape coefficients are equally treated as adjustable parameters of the model.
To describe accurately the measured spectrum, the model function in Eq. (\ref{SqE}) must be convoluted with the instrument resolution function $R(Q,E)$ and a spectral background should also be added to such a convolution. Explicitly:

\begin{equation}
\tilde{S}(Q,E)= R(Q,E)\otimes S(Q,E)+B(E),
\label{Iqw}
\end{equation}

\noindent where $B(E)$  is a typically mildly $E$-dependent background intensity. A Bayesian inferential procedure can be applied to determine, based on the measurements and on our prior knowledge of the physical problem under investigation, the most plausible shape, intensity and background parameters entering in Eq. (\ref{Iqw}) through Eqs. (\ref{SqE}) and (\ref{DHO}).

Best-fitting model lineshapes were determined using a Bayesian inferential analysis implemented through a Markov Chain Monte Carlo (MCMC) \cite{MarkovChain} routine with reversible jump (RJ) \cite{green1995reversible} steps. This approach can be used to probabilistically infer, based on a given measurement, the joint posterior probability distribution, or, in short, the posterior, of each parameter of the model. Notice that in the present case a prior uniform distribution for the number $k$ of inelastic modes has been deliberately chosen. In this respect the Bayesian procedure can be used to probabilistically ``rate'' a guess on the number of excitations in the sample scattering signal. In general, the knowledge of the entire posterior distribution of each  parameter also authorizes the interpretation of the maximum of such a posterior as the optimal (most plausible, or best-fitting) value of such a parameter. This assignment is possible provided the posterior, albeit not necessarily symmetric, is sharply peaked, well-shaped and unimodal. The shape itself of the posterior carries important information on the precision and the likelihood of a given best-fitting value. Foundations and working principle of the performed Bayesian analysis are discussed in great detail in Ref. [\citenum{de2016bayesian,de2019model,DefrancescoBook2020,de2019review}].

\section{Comparing measured and model spectral shapes }\label{Res}

\begin{figure}[H]
	\centering
	\includegraphics[width=1\linewidth]{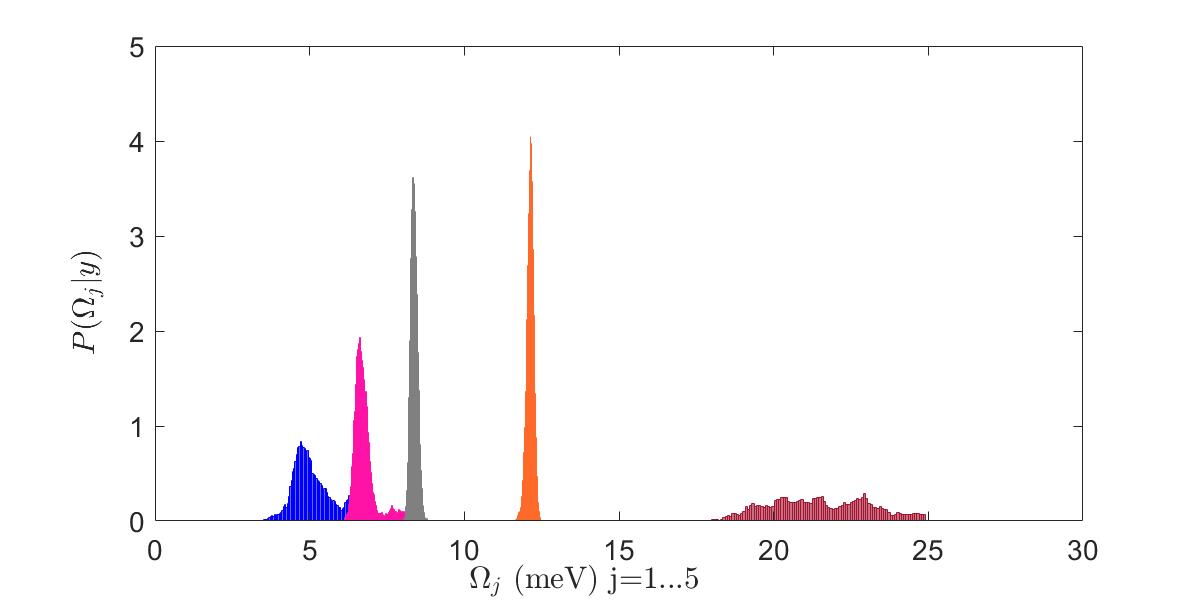}
	\caption{\textbf{Posterior distribution functions for the phonon mode frequencies in polycrystalline ice. } The plot displays the posterior distribution functions conditional on the experimental data $y$ for the phonon mode frequency (phonon peak position) in the spectrum of polycrystalline ice at 225 K and wavevector transfer $Q =$ 21 nm$^{-1}$ as obtained by the MCMC-RJ algorithm (see main text). Color code is as in the main text.}
	\label{Fig1S}
\end{figure}

\begin{figure}[H]
	\centering
	\includegraphics[width=1\linewidth]{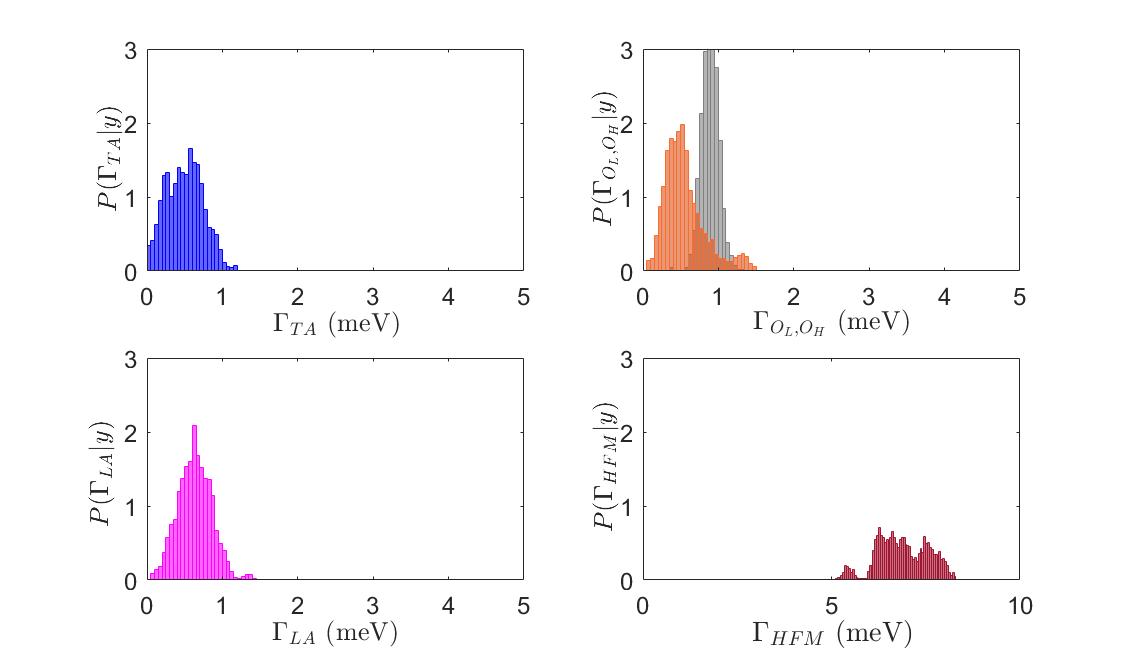}
	\caption{\textbf{Posterior distribution function for the collective excitations dampings in pure polycrystalline ice.} In the four panels we report the posterior distribution drawn based on experimental data $y$ for the damping coefficient of the five phonon modes identified in the spectrum of pure ice at $Q =$ 21 nm$^{-1}$.}
	\label{Fig2S}
\end{figure}

The posterior distributions delivered by the MCMC-RJ algorithm (see main text) for energy shifts and damping coefficients of the Damped Harmonic Oscillator (DHO) model components are displayed in Figs. \ref{Fig1S} and \ref{Fig2S}, respectively, as an example at $Q$= 21 nm$^{-1}$. The plots demonstrate how the Bayesian approach can draw the entire probability distribution for all model parameters, rather than providing best-fitting values only, as in a more traditional $\chi^2$-minimization approach. 
From Fig. \ref{Fig1S}, it readily appears that the majority of posteriors drawn for the DHO energy shifts are very sharp, unimodal, and symmetric about their maxima, their mode thus yielding a straightforward and unambiguous estimate of the best-fitting parameter value.
However, as illustrated in Fig. \ref{Fig2S}, the scenario is not always as rosy for the damping coefficients, i.e., the width of the DHO model profiles; this partially stems from the circumstance that is usually more challenging to determine reliably the widths than the corresponding peak positions.
In general, the width of the posterior distribution provides a direct measure of the uncertainty affecting the estimate of an optimal parameter value. This uncertainty can be secondary to the poor statistical accuracy of data, or can reveal a trend physically significant. For instance, a large posterior width for a mode frequency can indicate a gradual approach to a mode overdamping regime. 

To provide a meaningful example, we can consider the $Q$ increasing damping of the HFM, illustrated in Fig. S4.  As discussed in the main manuscript in further detail, such a high-energy phonon gradually damps off as $Q$ increases. This trend is also mirrored by the shape of the posterior distribution of both frequency and damping of such mode, as clearly emerges from considering the posteriors drawn for the highest energy phonon mode in Figs. \ref{Fig1S} and \ref{Fig2S}.  These relatively unstructured posteriors hamper a precise determination of the optimal parameter value. In the specific case, this  difficulty is exacerbated by the finite energy window of the available experimental data, which in some cases prevented the measurement to fully capture the position of the tails of the highest energy phonon mode.

\begin{figure}[H]
	\centering
	\includegraphics[width=1\linewidth]{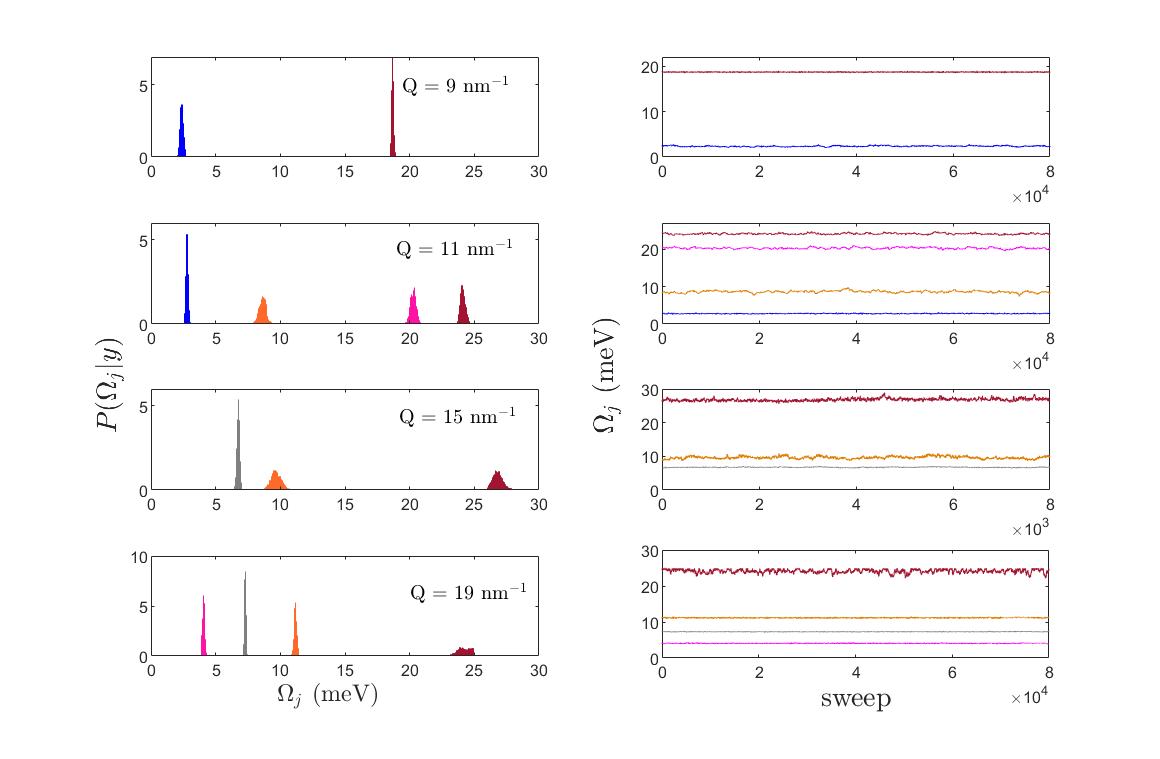}
	\caption{\textbf{Posterior distribution function for the excitation frequencies modes in ice at $T$ = 225 K}. In the left column are reported the posterior distribution functions for the excitation frequencies modes revealed on ice at four selected $Q$ values. The blue, grey, orange, magenta and wine colors indicate the distributions for the transverse mode, the low frequency optical mode, the high frequency optical mode, the longitudinal center of mass acoustic mode and the oxygen high frequency mode, respectively. In the right column the corresponding traceplots are reported with the same color code.}
	\label{Fig3S}
\end{figure}

\begin{figure}[H]
	\centering
	\includegraphics[width=1\linewidth]{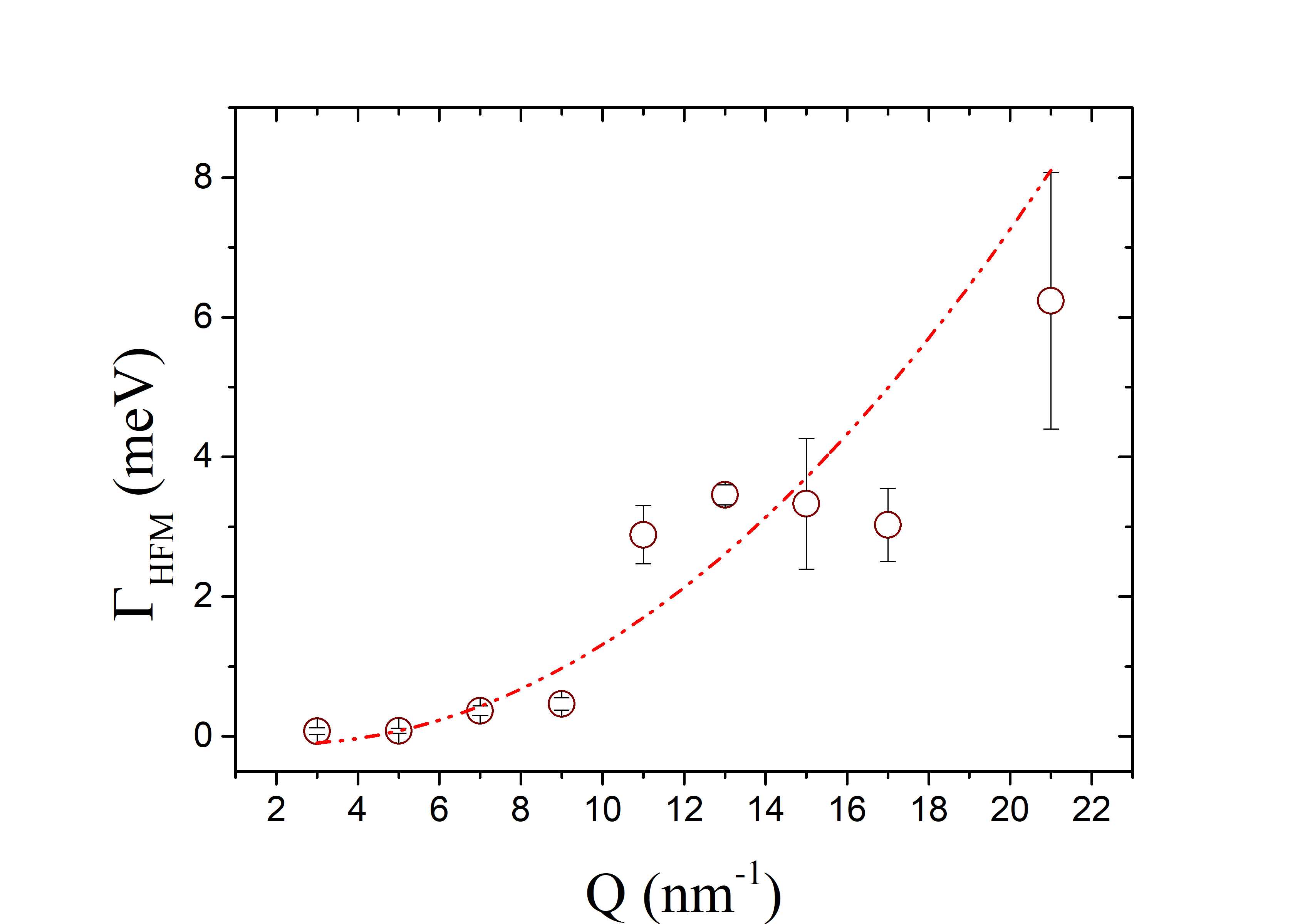}
	\caption{\textbf{Damping of the highest energy mode in pure ice at $T$ = 225 K}. Damping of the HFM associated to the collective excitations of oxygen atoms as a function of the wavevector transfer $Q$.}
	\label{Fig4S}
\end{figure}

In Fig. \ref{Fig3S} we give a graphical demonstration of how, depending on the value of the wavevector $Q$, the number of modes that it is possible to recognize in the IXS spectra changes as well as the greater or lesser ``easiness" with which they are solved. According to their dispersion curve and intensity, some modes may come out from the resolution, be clearly visible, succumb to more intense ones, or, finally, move out the investigated energy window. As in Fig. \ref{Fig1S}, in the left column of the figure we report the fairly well-shaped posteriors of the excitation frequencies of the different modes. In the right column, we show the corresponding traceplots for the revealed mode frequencies. These traceplots are one of the possible graphical (and diagnostic) methods MCMC users adopt to decide with sound theoretical support when the algorithm reaches convergence. The appearance of traceplots provides a clear visualization of the efficiency of the Markov Chain in exploring the parameter space. Bayesian statisticians usually call the correct look of a traceplot a \textit{hairy caterpillar}.

\begin{figure}[H]
	\centering
	\includegraphics[width=1\linewidth]{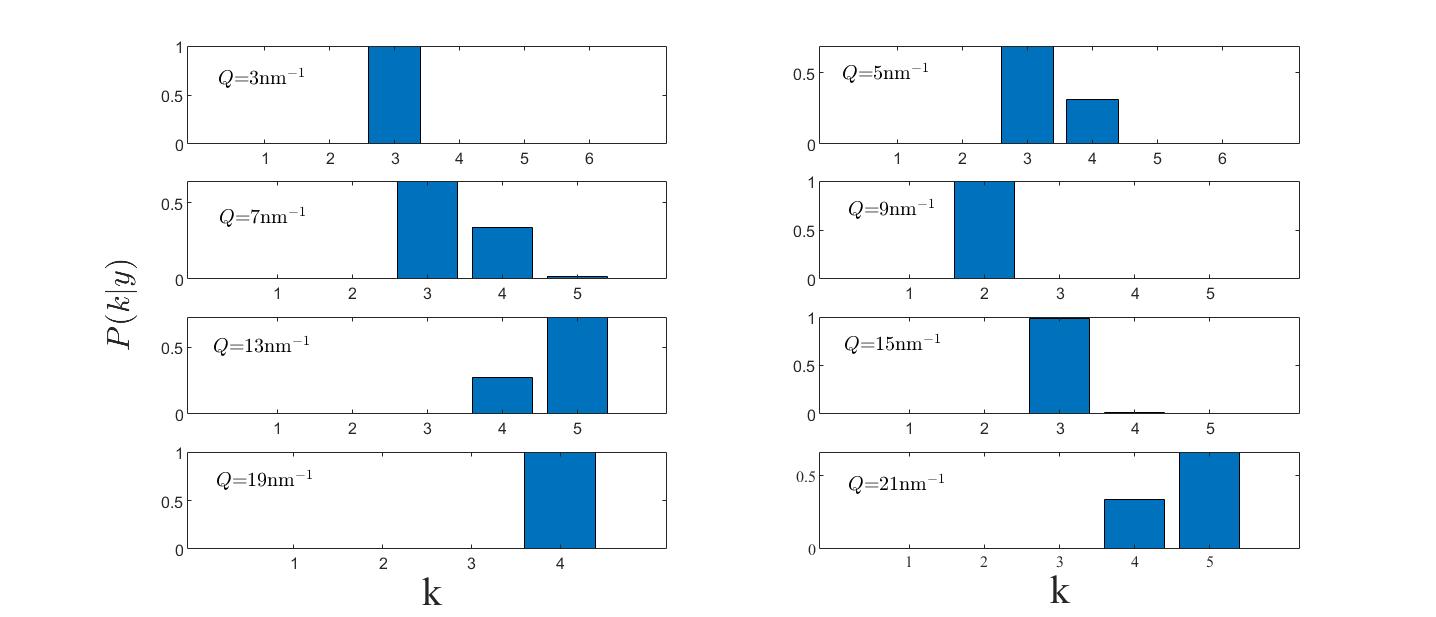}
	\caption{\textbf{Posterior distribution function for the number of modes in ice at selected momentum transfer values}. Posterior distribution for the number of inelastic component $k$ in the spectra of polycrystalline ice at the indicated momentum transfer Q conditional to the experimental data $y$. The values of such probabilities are reported also in the Table here below, in fact too small probabilities are not detectable in the histograms.}
	\label{Fig5S}
\end{figure}

To provide an example of the uniqueness and the soundness of the model description within the Bayes inference framework, in Fig. \ref{Fig5S}, we report the posterior probability distribution of the number of excitations, conditional to the measurements obtained at different $Q$'s. This distribution is represented by the histograms drawn by the Bayesian algorithm. The distribution mode, i.e., the position of the tallest histogram rectangle, identifies the optimal parameter value. 
Adopting a model with such an optimal value might appear a non-unequivocal choice if competing suboptimal model options have comparable probability; however, such a choice is the one probabilistically most grounded. 
Notice that suboptimal solutions having a too small probability are not readily visible in the histograms, for this reason, all probabilities are also reported in the numerical table here below.

\begin{table}[H]
\caption{\label{tab:table1}
Posterior probabilities for the number of inelastic component in the spectra at the indicated different $Q$ values conditional to the experimental data $y$}
\begin{ruledtabular}
\begin{tabular}{cccccc}
 \\ &\multicolumn{5}{c}{$k$}\\ \cline{2-6}\\
 Q (nm$^{ -1}$) & 1 & 2 & 3 & 4 & 5\\
\hline\\
3 & 0 & 0 & 0.9995& 0.0005 & 0 \\
5 & 0 & 0 & 0.6883 &  0.3095 & 0.0022\\
7 & 0 & 0 & 0.6413 & 0.3401 & 0.0186 \\
9 & 0 & 0.9981  & 0.0019 & 0 & 0 \\
11& 0 & 0 & 0.9225 & 0.0775\\   
13& 0 & 0 & 0 & 0.2741 & 0.7259 \\
15& 0 & 0 & 0.9832  & 0.0168 & 0 \\
17& 0 & 0 & 0.9382 & 0.0612 & 0.0006\\  
19& 0 & 0 & 0 & 1 & 0 \\
21& 0 & 0 & 0 & 0.3381 & 0.6619 \\

\end{tabular}
\end{ruledtabular}

\end{table}


%\bibliography{your-bib-file}

 %``” 
%\bibliographystyle{plain}
%\bibliography{biblio_ice}

\end{document}